# Relay Assisted Multiuser OWC Systems under Human Blockage

Y. Zeng[1], Sanaa H. Mohamed[2], Ahmad Qidan[2], Taisir E. H. El-Gorashi[2], Jaafar M. H. Elmirghani[2]

[1]School of Electronic and Electrical Engineering, University of Leeds, LS2 9JT, United Kingdom
[2]Department of Engineering, King's College London, WC2R 2LS, United Kingdom
e-mail: ml16y5z@leeds.ac.uk.

**ABSTRACT**

This paper proposes using cooperative communication based on optoelectronic (O-E-O) amplify-and-forward relay terminals to reduce the influence of the blockage and shadowing resulting from human movement in a beam steering Optical Wireless Communication (OWC) system. The simulation results indicate that on average, the outage probability of the cooperative communication mode with O-E-O relay terminals is two orders of magnitude lower than the outage probability of the system without relay terminals.

## 1. INTRODUCTION

According to the analysis of CISCO, Internet users are predicted to grow from 3.9 billion users in 2018 to 5.3 billion users by 2023 [1]. Various technologies are advocated and investigated for providing ultra-fast wireless communication for customers. Among these is optical wireless communication (OWC) which received extensive attention from research communities due to its high efficiency, low start-up cost, and low operational costs. However, considering the optical propagation property of InfraRed (IR) and Visible Light Communication (VLC) signal used in OWC systems, the OWC signal is more susceptible to obstruction compared conventional Radio Frequency Communication (RFC) systems [2]-[3]. The blockage and shadowing impact due to stationary furniture is predictable and can be evaded by rational transmitter location setup. Conversely, humans randomly move in the room with low speed and their real time locations of humans are unpredictable by OWC systems.

One of the main methods used to address blockage and shadowing due to moving humans is multiple signal sources collaboration. The blockage experienced by a rotating user in an environment with fixed obstacles is investigated [4], [5]. Fixing the location of obstacles simplifies the analysis.it. The blockage and shadowing of moving obstacles is simplified as a parameter of a Bernoulli distribution to investigate the use of cooperating multiple Access Points (APs) to improve the performance of the OWC system [6]. In [7], the author introduced a Random Way Point (RWP) model to simulate human movement in the room where the human is modelled as a cylinder to evaluate the blockage and shadowing. The results revealed that an OWC system with 8 APs have better performance than a system with 4 APs. RWP models are also used to model humans to study the performance of OWC systems under blockage considering a transmitter with a Lambertian model [8]. Multi APs or transmitters cooperation was considered as the solution of human blockage in [4-8]. However, the above papers focused on the diffuse transmitters. This setup faces violent ambient light due to the wide FOV of the receiver and multipath dispersion due to the reflection of optical beams at walls and ceiling.

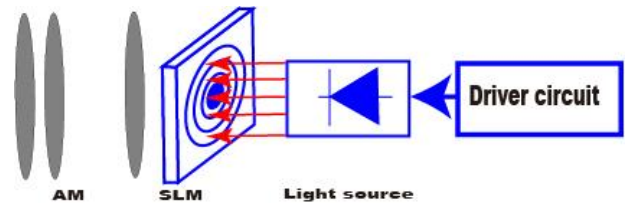

*Figure 1 The beam steering configuration of the transmitters.*

Beam steering was proposed in recent studies to solve the drawbacks of diffuse transmitters [9]-[11]. The narrow beam size reduces free space propagation loss and the Inter Symbol Interference (ISI) due to the reflection from walls and ceiling [12]-[14]. The authors in [10] used beam steering to achieve a data rate of up to 100 Gbps. Fig. 1 shows the structure of a typical beam steering transmitter for OWC systems. A digital Fresnel Zone Plate (FZP) pattern is rendered on a spatial light modulator [9]. With changing the zone center of the FZP, the parallel light beams from the light source are focused on the far field with arbitrary steering angle [10]-[13]. However, the maximum beam angle is limited by the pixel size of SLM. An Angle Magnifier (AM), constructed from 3 lens (shown in Fig.1), is proposed in [11] to extend the maximum beam steering angle of the transmitter. A beam steering transmitter with maximum beam steering angle at 30 degrees was reported in [9]. In this paper, we introduced relay terminals to solve the problem of human blockage in a beam steering OWC system. The performance of relay terminals in a multiuser beam steering based OWC system is investigated in terms of outage probability in a simulation environment based on the ray tracing algorithm used in [12] and [13].

The rest of the paper is organized as follows: The direct and cooperative communication modes for multiuser beam steering OWC system are introduced in Section 2. The simulation configuration and the results are presented in Section 3. Finally, the conclusions are presented in Section 4.

## 2. THE DIRECT AND COOPERATIVE COMMUNICATION MODES FOR MULTIUSER BEAM STEERING OWC SYSTEM

We consider a realistic indoor environment where multiple APs are mounted on the ceiling as shown in Fig. 2. All APs are connected with fiber and are connected to a central unit that controls the resources of the network and allocates them fairly among the users. Multiple users were distributed in the room. We assume an AP finds the location of the users based on Location Estimation Algorithm (LEA) proposed in [14] and acquire perfect Channel State Information (CSI) of relevant users. Using Non-Orthogonal Multiple Access (NOMA) and unipolar intensity modulation (LDs only works in positive area), each AP transmits real and positive signals to intended users. The signal from each AP is a superposition of the signals that convey the information for intended users with different power levels based on the CSI of different users [15] and [16].

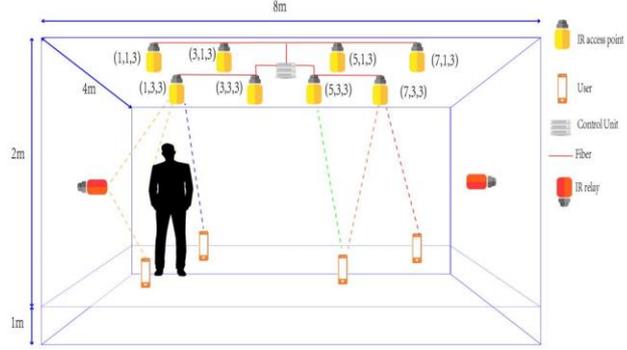

*Figure 2 Human body block the signal from the Access.*

Two different communication modes are considered in studying the influence of human blockage. The first mode is the direct communication mode. APs directly sends signal to users. The second mode is the cooperative communication mode based on relay terminals. In the first phase of r the cooperative communication mode, APs sends signals to users and relay terminals in the vicinity of it. Optoelectronic (O-E-O) amplify-and-forward relay terminals are considered in this paper. The optical signal from the AP is transformed into an electrical signal via a PIN photodetector. Then, the electrical signal as well as the noise are amplified by an electrical amplifier before modulating an IR laser. We assume that the parameters of the O-E-O amplify-and-forward relay terminal are constant within the noise equivalent bandwidth. In the second phase, the amplified signal is forwarded to the user. Maximum Ratio Combining (MRC) is used at the user's device to combine the received signal from both phases [17]-[20].

For the direct communication mode, considering the influence of the blockage and shadowing, the Signal to Interference and Noise Ratio (SINR) is defined as [21] and [22]:

$$SINR_i = \frac{\sum_{l=1}^{L}(\beta_{li}P_{li}R_iH_{li})^2}{\sigma_i^2 + \sum_{l=1}^{L}\sum_{k>i}(\beta_{li}P_{lk}R_iH_{li})^2}, \quad (1)$$

where $L$ is the total number of APs, $R_i$ is the responsivity of user $i$, $\sigma_i^2$ is the additive Gaussian noise at user $i$, $P_{li}$ is the optical power received by user $i$ from AP $l$, $H_{li}$ is the CSI between user $i$ and AP $l$ and $\beta_{li}$ is the blockage factor for the link between AP $l$ and user $i$, $\beta_{li} = 1$ if the link is not blocked and $\beta_{li} = 0$ if the link is blocked.

For the cooperative communication mode, the output signals are combined using MRC through an adder circuit. Each input to the circuit is directly proportional to its SNR by using a relevant weight. The SINR of user $i$ based on cooperative communication is given as:

$$SINR_i^{MRC} = SINR_i^{first\ phase} + SINR_i^{second\ phase}$$
$$= \frac{\sum_{l=1}^{L}(\beta_{li}P_{li}R_iH_{li})^2}{\sigma_i^2 + \sum_{l=1}^{L}\sum_{k>i}(\beta_{li}P_{lk}R_iH_{li})^2} + \frac{\sum_{r\in R}(\gamma_{lri}P_{li}R_iH_{lr}H_{ri})^2}{\sigma_i^2 + \sum_{k>i}(\gamma_{lri}I_{lk}R_iH_{lr}H_{ri})^2 + \sum_{r\in R}\gamma_{lri}\sigma_r^2}, \quad (2)$$

where $\gamma_{lri}$ is the blockage factor for the link between AP $l$ and user $i$ via relay $r$, $\gamma_{lri} = 1$ if the link is not blocked and $\gamma_{lri} = 0$ if the link is blocked. The impact of the human body blockage is evaluated by studying the outage probability which is defined as the following:

$$P_{out} = P(SNR_i \leq SNR_{th}), \quad (3)$$

where $SNR_{th}$ is the SNR threshold (15.6 dB) for normal communication which is equivalent to a bit error rate of $10^{-9}$ [23] and [24].

## 3. SIMULATION SETUP AND RESULTS

### 3.1 Simulation Setup

In order to evaluate the advantages of the proposed method, a simulation was performed considering an empty room with dimensions of 8 m × 4 m × 3 m (length × width × height). The plaster walls are assumed to reflect light rays in a form close to a Lambertian function. Therefore, the walls (including the ceiling) and floor were modelled as Lambertian reflectors with reflectivity of 80% and 30%, respectively. Reflections from doors and windows are identical to reflections from walls. The transmitted signals are reflected from the room reflecting surfaces. The reflecting surfaces are divided into a number of equal-sized, square-shaped reflection elements. The reflection elements have been treated as small transmitters that diffuse the received signals from their centers in a Lambertian pattern [18]-[19]. It is noted that third-order reflections and higher do not produce a significant change in the received optical power, and therefore reflections up to second order only are considered [21]-[23]. The surface element size used in this work is set to 5cm × 5cm for the first-order reflections and 20cm × 20cm for the second-order reflections. The illumination system consists of 8 LDs as in [19].

The locations of the APs are shown as Figure 2. We consider six users in the room in the locations depicted in Table 1. The locations of the users are set to represent different situations where the users are served by 1, 2 or 4 APs. Eight relay terminals mounted on the walls. Each of them can receive the signal from one AP. The rest of the parameters are shown in Table 1.

*Table 1. Simulation parameters*

| Parameter | Configuration | |
|---|---|---|
| Room set | | |
| Length, Width, Height | 8m, 4m, 3m | |
| Reflectivity of walls | 0.8 | |
| Reflectivity of ceiling | 0.8 | |
| Reflectivity of floor | 0.3 | |
| IR AP | | |
| Quantity | 8 | |
| Location | (1,1,3) (1,3,3) (1,5,3) (1,7,3) (3,1,3) (3,3,3) (3,5,3) (3,7,3) | |
| The average optical power | 1mW | |
| beam divergency | 2.1mrad | |
| IR relay terminal | | |
| Location | (0m,1m,1.5m), (0,3,1.5), (0,5,1.5), (0,7,1.5), (4,1,1.5), (4,3,1.5), (4,5,1.5) and (4,7,1.5) | |
| Users | | |
| Quantity | 6 | |
| Location | (1m ,1m,1m), (1,4,1), (1,7,1), (2,1,1), (2,4,1) and (2,7,1) | |
| Elevation | 90º | |
| Azimuth | 0º | |
| Area | 1cm$^2$ | |
| The responsivity of IR | 0.5A/W | |
| The responsivity of illumination | 0.4A/W | |
| Field Of View (FOV) | 90º | |
| Resolution | | |
| Time bin duration | 0.01 ns | |
| Bounces | 1 | 2 |
| Surface elements | 32000 | 2000 |
| Wavelength | 850nm | |
| Bandwidth (BW) | 10 GHz | |

To simulate the blockage and shadowing due to moving humans, we consider the RWP model, the human moves along a straight line at a uniform velocity to a random destination in the room. Before reaching its destination, the, human selects another destination randomly according to a uniform distribution and moves toward it at a different velocity. For the sake of simplicity and without loss of generality, we use the RWP model without any pause time at any destination. As the human continues moving in the room, the probability density function $P_{pdf}$ of this human location in a rectangle room is defined as follows [6]:

$$P_{pdf} = f(x)f(y) = \frac{36}{x_m^3 y_m^3}\left(x^2 - \frac{x_m^2}{4}\right)\left(y^2 - \frac{y_m^2}{4}\right), -\frac{X_m}{2} \leq x \leq \frac{X_m}{2}, -\frac{Y_m}{2} \leq y \leq \frac{Y_m}{2}, \quad (4)$$

where $(x, y)$ is the location of this human on the floor, $(x_m, y_m)$ is the size of the floor in two dimensions. A human is modelled as a cylinder (Height=1.8m, Radius=0.3m). For a link between transmitter $(x_t, y_t, z_t)$ and receiver $(x_r, y_r, z_r)$, where the transmitter can be an AP or a relay terminal, and the receiver can be a relay terminal or a user, we can find an area $S$ in the $(x, y)$ plane where the human (cylinder) block the direct link between the transmitter and the receiver. Then we can get the cumulative probability density function $P_{l_{tr}}$ for the block factor (e.g. $\beta_{li}$ or $\gamma_{lri}$) of the LOS link $l_{tr}$ between the transmitter and the receiver by the following equation:

$$P_{l_{tr}} = P_{cdf} = \iint_S f(x)f(y) \quad (5)$$

### 3.2 Simulation Results

The results of the outage probability of the two communication modes are shown in Fig. 3. The outage probability of cooperative communication mode is lower than direct communication mode for all users. For user 5, the outage probability of the cooperative communication mode is almost three orders of magnitude lower than the direct communication mode. User 5 is located at the center of the room. The outage probability of direct communication mode of user 5 is higher than other users. With cooperative communication, user 5 can receive signals form 4 relay terminals which efficiently reduce the influence of the blockage. Compared with user 5, the outage probability degradations of the other users is distributed in the range from one order of magnitude to two orders of magnitude. The degradation degree is related to the number of relay nodes the user can receive signal from.

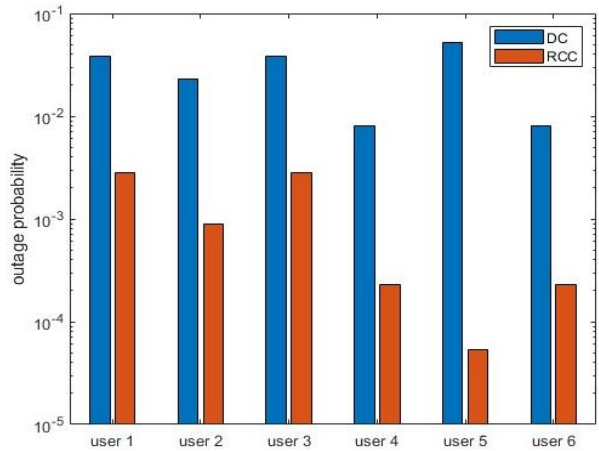

*Figure 3 THE OUTAGE PROBABILITY OF DIRECT COMMUNICATION MODE AND RELAY COOPERATIVE COMMUNICATION MODE FOR SENARIO 1.*

### 4.CONCLUSION

In this paper, we proposed using cooperative communication based on optoelectronic (O-E-O) amplify-and-forward relay terminals to reduce the influence of the blockage and shadowing resulting from human movement in a beam steering Optical Wireless Communication (OWC) system. The cooperative communication system consists of two phases. In the first phase, APs transmit the signal to relay terminals and user. In second phase, relay terminals forward received signal to users. The user combines the signals of the two phases using MRC. The simulation results show that outage probability of the cooperative communication mode is two orders of magnitude lower on average than direct communication without relay terminals. The results show that the user that has links with more relay terminals observes higher robustness to the human body blockage.

### ACKNOWLEDGMENTS
The authors would like to acknowledge funding from the Engineering and Physical Sciences Research Council (EPSRC) TOWS (EP/S016570/1) project.